\documentclass[aps,prl,twoside,twocolumn,floatfix,a4paper,showpacs]{revtex4-1}
\usepackage{amsmath}
\usepackage{amsfonts}
\usepackage{amssymb}
\usepackage{graphicx,epsfig}
\usepackage{placeins}
\usepackage{pbox}
\usepackage{color}

\newcommand{\la}{\left\langle}
\newcommand{\ra}{\right\rangle}

\begin{document}

\title{Wiener-Khinchin theorem for nonstationary scale-invariant processes}

\author{Andreas Dechant and Eric Lutz}
\affiliation{Department of Physics, Friedrich-Alexander Universit\"at Erlangen-N{\"u}rnberg, D-91058 Erlangen, Germany}

\begin{abstract}
We derive a generalization of the Wiener-Khinchin theorem for nonstationary processes by introducing  a time-dependent spectral density that is related to the time-averaged power. We use the nonstationary theorem to investigate aging processes with  asymptotically scale-invariant correlation functions.  As an application, we analyze the power spectrum of three paradigmatic models of anomalous diffusion: scaled Brownian motion, fractional Brownian motion and diffusion in a logarithmic potential. We moreover elucidate  how the nonstationarity of generic subdiffusive  processes is related to the infrared catastrophe of $1/f$-noise.
\end{abstract}

\maketitle

The Wiener-Khinchin theorem  is a fundamental result of  the theory of stochastic processes. In its simplest form, it states that the  autocorrelation function, $C(\tau) = \la x(t+\tau) x(t)\ra$, of a stationary random signal $x(t)$ is equal to the Fourier transform of the power spectral density, $S(\omega) =\la |\hat x(\omega)|^2\ra/T$ \cite{wie30,khi34}. Here $\hat x(\omega)$ denotes the Fourier transform of $x(t)$, $T$ the total measurement time and the ensemble average $\la...\ra$ is taken over many realizations of the signal. The Wiener-Khinchin theorem provides a simple relationship between  time and frequency representations of a fluctuating process.  Its great practical importance comes from the fact that the autocorrelation function may be directly determined from the measured  power spectral density, and vice versa. Over the years, it has become an indispensable tool in signal and communication theory \cite{zie14}, electric engineering \cite{yat04}, the theory of Brownian motion \cite{ris89}, classical \cite{goo85} and quantum \cite{man95} optics,  to name a few. 

The Wiener-Khinchin theorem is only applicable to weakly stationary processes that are characterized by a time-independent mean $\la x(t)\ra$ and a correlation function $C(\tau)$ that solely depends on the difference $\tau$ of its time arguments \cite{coh95}. It fails for nonstationary processes with an autocorrelation function, $C(\tau,t) = \langle x(t+\tau) x(t) \rangle$, that depends explicitly on  time $t$ and hence exhibits aging. Further, the power spectral  density, $S(\omega,t)$---now an explicit function of both frequency and time---is not uniquely defined for nonstationary noisy signals \cite{coh95}. A commonly used generalization, the Wigner-Ville function \cite{wig32,vil48}, given in Eq.~\eqref{wigner-ville} below, is related to the instantaneous power; it  has the disadvantage of not being a true spectral density as it can take on negative values \cite{coh95}. We here address these critical issues by first introducing a  spectral  density that is related to the time-averaged power. Contrary to the Wigner-Ville function, it is strictly positive and therefore a proper spectral density. We employ this quantity to derive an extension of the Wiener-Khinchin theorem that is valid for finite-time, nonstationary signals. We apply this generalization to aging processes that are characterized by a correlation function of the form, $C(\tau,t) \simeq \mathcal{C} t^{\alpha} \phi({\tau}/{t}) $, where $\phi$ is an arbitrary scaling function. Correlation functions of this type describe scale-invariant dynamics that do not possess a distinctive time scale, contrary to exponentially correlated processes. They occur in a wide range of physical \cite{bou90}, chemical \cite{ben00} and biological \cite{wes94} systems and have also found applications in  finance \cite{man97} and the social sciences \cite{bro06}. We illustrate our results by analyzing three important anomalous diffusion processes \cite{met00}: scaled Brownian motion, fractional Brownian motion and diffusion in a logarithmic potential. We obtain for each of them analytical expressions for the  nonstationary  spectral density in the low/high frequency domain, and  demonstrate the usefulness   of the generalized Wiener-Khinchin theorem as a tool to perform spectral analysis in the aging regime. We finally discuss the general connection between subdiffusion  and $1/f$ noise \cite{dut81}.

\textit{Nonstationary Wiener-Khinchin theorem.} Let us begin by defining our notation. We consider a continuous random signal $x(t)$ which is measured on the time interval $t\in[0,T]$. As customary, we introduce the  truncated Fourier transform of  its time-shifted version $y(\theta) = x(\theta+T/2)$ on $\theta \in [-T/2,T/2]$ \cite{zie14}:

\begin{align}
\hat{y}(\omega,T) =  \int_{-T/2}^{T/2} \text{d}\theta \ e^{-i \omega \theta} y(\theta) \label{truncated-fourier}.
\end{align}
 Equation \eqref{truncated-fourier} reduces to the usual Fourier transform in the limit $T$ to infinity. 
The  spectral density $S(\omega)$ is  defined through the ensemble average of the modulus squared of $\hat{y}(\omega,T)$ \cite{zie14},
\begin{align}
S(\omega) &= \lim_{T \rightarrow \infty} \frac{1}{T} \langle |\hat{y}(\omega,T) |^2 \rangle  \label{spectral-definition-stationary}.
\end{align}
Note that Eq.~\eqref{spectral-definition-stationary} has to be interpreted in the distributional sense for nonintegrable functions.
In the following, we assume that the stochastic process $x(t)$ is  real. As a result, the power spectrum $S(\omega)$ and the stationary correlation function $C(\tau)$ are even functions.
Physically, the  spectral density  measures the power contained in the signal at a certain frequency \cite{zie14}. 
More precisely, $\int_{\omega_1}^{\omega_2} \text{d}\omega \ S(\omega)/\pi $ corresponds to  the power contained in the frequency interval $[\omega_1,\omega_2]$ and the total power $\langle x^2 \rangle$ is obtained by integrating over all frequencies, $\langle x^2 \rangle = \int_{-\infty}^{\infty} \text{d}\omega \ S(\omega)/({2}\pi)$.
According to  the Wiener-Khinchin theorem  $S(\omega)$ and $C(\tau)$ are related via \cite{wie30,khi34},
\begin{align}
S(\omega) &= \int_{-\infty}^{\infty} \text{d}\tau  e^{-i \omega \tau} C(\tau) = 2\int_0^\infty d\tau  \cos(\omega \tau) C(\tau)\label{wiener-Khinchin}.
\end{align}
Equations \eqref{spectral-definition-stationary} and \eqref{wiener-Khinchin} may be regarded as two definitions of the power spectrum. They are fully equivalent for stationary processes \cite{coh95}.

For nonstationary random signals, Eqs.~\eqref{spectral-definition-stationary} and \eqref{wiener-Khinchin} are no longer equivalent and both expressions may be used to define a nonstationary generalization of  the power spectrum.
Starting from Eq.~\eqref{wiener-Khinchin}, the Wigner-Ville spectrum $S_\text{wv}(\omega,t)$ is constructed as the Fourier transform of the symmetrized correlation function \cite{mar85},
\begin{align}
S_\text{wv}(\omega,t) = \int_{-\infty}^{\infty} \text{d}\tau \ e^{-i \omega \tau} \Big\langle x\Big(t+\frac{\tau}{2}\Big) x\Big(t-\frac{\tau}{2}\Big) \Big\rangle . \label{wigner-ville}
\end{align}
Equation \eqref{wigner-ville} is simply related to the nonstationary correlation function, $C(\tau,t)= \int_{-\infty}^\infty (d\omega/2\pi) e^{i\omega \tau} S_\text{wv}(\omega,t)$, thus preserving the usual form of the Wiener-Khinchin theorem. However, it is not a proper   spectral density since it can take on negative values. Another drawback is that it depends on the infinitely extended past and future of a process. This does not pose a problem if the investigated signal is non-zero only in a finite time interval. But the truncation of the correlation function of an extended signal which results from a finite measurement time will lead to deviations from the Wigner-Ville spectrum that are not taken into account in Eq.~\eqref{wigner-ville}.

We here  follow a different path and interpret Eq.~\eqref{spectral-definition-stationary} as the definition of the spectral density. Thus, for a finite measurement time $T$, we introduce the quantity,
\begin{align}
S_\text{ta}&(\omega,T) = \frac{1}{T}\langle |\hat{y}(\omega,T) |^2 \rangle \nonumber \\
& = \frac{1}{T} \int_{-T/2}^{T/2} \text{d}\theta' \int_{-T/2}^{T/2} \text{d}\theta'' \ e^{i \omega(\theta'-\theta'')} \langle y(\theta') y(\theta'') \rangle . \label{5}
\end{align}
This definition ensures that $S_\text{ta}(\omega,t)$ is positive and hence a true  spectral density, contrary to the Wigner-Ville spectrum  \eqref{wigner-ville}.
For stationary systems, we have, 
\begin{align}
S(\omega) = \lim_{T \rightarrow \infty} S_\text{ta}(\omega, T) = S_{\text{wv}}(\omega,0),
\end{align}
so that all three definitions of the spectral density coincide in the long-time limit.
For nonstationary processes, $S(\omega)$ does not exist, whereas $S_\text{ta}(\omega,T)$ and $S_\text{wv}(\omega,t)$ are generally distinct.
The Wigner-Ville spectrum  $S_\text{wv}(\omega,t)$ is related to the instantaneous power, while $S_\text{ta}(\omega,T)$ corresponds to the time-averaged (ta) power,
\begin{align}
\frac{1}{2 \pi} \int_{-\infty}^{\infty} \text{d}\omega \ S_\text{wv}(\omega,t) &= \langle x^2(t) \rangle, \label{spectrum-wv} \\ 
\frac{1}{2 \pi} \int_{-\infty}^{\infty} \text{d}\omega \ S_\text{ta}(\omega,T) &=\frac{1}{T} \int_{0}^{T} \text{d}t' \ \langle x^2(t')\rangle. \label{spectrum-power}
\end{align}
We note that Eqs.~\eqref{spectrum-wv} and \eqref{spectrum-power} differ for nonstationary signals.
Equation \eqref{spectrum-power}  provides a clear physical interpretation of $S_\text{ta}(\omega,T)$: $\int_{\omega_1}^{\omega_2} \text{d}\omega \ S_\text{ta}(\omega,T)/\pi$ is the time-averaged power contained in the frequency interval $[\omega_1,\omega_2]$, in complete analogy to the stationary case. 

From the definition \eqref{5}, we find that the time-averaged spectral density $S_\text{ta}(\omega,T)$ can be expressed in 
 terms of the correlation function $C(\tau,t)$ as,
\begin{align}
S_\text{ta}(\omega,T) &= \frac{2}{T} \int_{0}^{T} \text{d}t'' \int_{0}^{t''} \text{d}t'  \cos(\omega t') C(t',t''-t'), \label{spectral-nonstationary}
\end{align}
where we have used that $C(\tau,t) = C(-\tau,t+\tau)$ so that both arguments of the correlation function are positive.
On the other hand, the autocorrelation function is related the spectral density $S_\text{ta}(\omega,t)$ via,
 \begin{align}
C(\tau,T-\tau) = \frac{1}{ \pi} \Big[1 + T \frac{\partial}{\partial T} \Big] \int_{0}^{\infty} \text{d}\omega  \cos(\omega \tau) S_\text{ta}(\omega,T), \label{spectral-nonstationary-inverse}
\end{align}
for $T > \tau$. Equations \eqref{spectral-nonstationary} and \eqref{spectral-nonstationary-inverse} constitute a nonstationary, finite-time generalization of the Wiener-Khinchin theorem \eqref{wiener-Khinchin}. It is free from the problems faced by the Wigner-Ville function \eqref{wigner-ville}; the  price to pay is that the familiar symmetry between time-frequency  representations  known from the Fourier transform is lost.  In the limit of stationary signals, Eqs.~\eqref{spectral-nonstationary} and \eqref{spectral-nonstationary-inverse} reduce to the usual Wiener-Khinchin theorem \eqref{wiener-Khinchin}.

\textit{Application to scale-invariant processes}. In order to illustrate the usefulness of the nonstationary Wiener-Khinchin relation \eqref{spectral-nonstationary}-\eqref{spectral-nonstationary-inverse}, we next consider aging processes with a correlation function of the scaling form \cite{dec14}, 
\begin{align}
C(t,\tau) \simeq \mathcal{C} t^{\alpha} \phi\left(\frac{\tau}{t}\right) . \label{scaling-correlation}
\end{align}
Here $\mathcal{C} > 0$ is a constant, $\alpha >-1$ the scaling exponent and $\phi(u)$  an arbitrary scaling function. Rescaling of the time variable only changes the prefactor of Eq.~\eqref{scaling-correlation} and not its qualitative behavior.
This kind of autocorrelation function is understood to describe the long-time  behavior of a system, where both the age $t$ and the time lag $\tau$ are large compared to the system's intrinsic time scales. In the absence of a characteristic correlation time, the only relevant time scale is the age $t$ of the system, which also governs the decay of the correlations as a function of the time lag $\tau$. Scaling correlation functions of this type appear in a large number of anomalous diffusion problems as we will discuss in the next section.

We may now use Eq.~\eqref{spectral-nonstationary} to compute the  time-averaged spectral density $S_\text{ta}(\omega,T)$ for the scaling correlation function \eqref{scaling-correlation}. We immediately find,
\begin{align}
&S_\text{ta}(\omega,T) \simeq 2 \mathcal{C} T^{\alpha+1} \chi(\omega T) \qquad \text{with} \nonumber \\
&\chi(z) = \int_{0}^{1} \text{d}v \ v^{\alpha+1} \cos(v z) \int_{\frac{v}{1-v}}^{\infty} \text{d}u \ u^{-\alpha-2} \phi(u) \label{scaling-wiener-Khinchin}.
\end{align}
Remarkably, the time-averaged spectral density has a scaling form,  similar to the autocorrelation function \eqref{scaling-correlation}, with a scaling function $\chi(\omega T)$.
Using Eq.~\eqref{spectral-nonstationary-inverse}, we further obtain the corresponding inverse relation,
\begin{align}
\phi(u) = \frac{1}{\pi} \int_{0}^{\infty} \text{d}z \ \cos({z u}) \big[(\alpha+2) \chi(z) + z \chi'(z) \big] \label{scaling-wk-inverse}.
\end{align}
The nonstationary  Wiener-Khinchin theorem \eqref{spectral-nonstationary}-\eqref{spectral-nonstationary-inverse} thus leads  to a direct relationship between the time and frequency scaling functions $\phi(u)$ and $\chi(z)$.
While the behavior of  $\chi(z)$ depends in a nontrivial manner on $\phi(u)$ and the  exponent $\alpha$, the mere existence of the scaling form \eqref{scaling-wiener-Khinchin}  already implies a number of generic  properties for the spectral density $S_\text{ta}(\omega,T)$.
We first note that the scaling function $\chi(\omega T)$ depends on the product of frequency and measurement time.
This means that low and high frequency limits of $S_\text{ta}(\omega,T)$ are intimately connected  to the measurement time $T$.
For low frequencies such that $\omega T \ll 1$ (or, equivalently, short  measurement times for a fixed frequency), Eq.~\eqref{scaling-wiener-Khinchin} may be easily evaluated since $\chi(0)$ is just a constant, and we obtain $S_\text{ta}(\omega,T) \simeq 2 \mathcal{C} T^{\alpha+1} \chi(0)$.
In  the low-frequency limit, the spectral density hence increases algebraically with the measurement time and is independent of frequency to leading order.
This reflects the fact that frequency components, whose period is much longer than the measurement time, will be essentially constant over the measurement interval.
For these frequencies, the truncated Fourier transform Eq.~\eqref{truncated-fourier} reduces to the time average $\bar{x}(T) = \int_{0}^{T} \text{d}t \ x(t)/T$ of the process $x(t)$.
Since by definition, $S_\text{ta}(0,T) = T \langle \bar{x}^2(T) \rangle$, we find that $\langle \bar{x}^2(T) \rangle \simeq 2 \mathcal{C} \chi(0) T^{\alpha-1}$.
This implies that processes described by the scaling correlation \eqref{scaling-correlation} are markedly different depending on the value of the  $\alpha$.
For $\alpha < 0$,  processes are ergodic, as the variance of the time average tends to zero, whereas for $\alpha > 0$ it increases with time  and processes are generally nonergodic \cite{pap91}.

In the high-frequency regime, $\omega T \gg 1$ (or large measurement times for a given frequency), we may explicitly  compute the asymptotic expansion of the function $\chi(z)$ for $z \gg 1$:
\begin{align}
\chi&(z) \simeq a_\lambda \bigg[ - \frac{\Gamma(\lambda+1)}{\alpha-\lambda+1} \sin\Big(\frac{\pi \lambda}{2}\Big) z^{-\lambda-1} \nonumber \\
& + \Gamma(\lambda+2) \cos\Big(\frac{\pi \lambda}{2}\Big) z^{-\lambda-2} \bigg] + \mathcal{O}(z^{-\lambda-3}) , \label{scaling-function-long-time}
\end{align}
provided that the scaling function asymptotically behaves as $\phi(u) \simeq a_\lambda u^{\lambda}$ with $\lambda > -1$ for $u \ll 1$ (see details in the Supplementary Material \cite{sm}).
We observe that the frequency dependence of the spectral density $S_\text{ta}(\omega,T)$ in the long-time limit depends on the small argument expansion of the scaling function $\phi(u)$ and, hence, on the long-time behavior of the correlation function for $t \gg \tau$.
If the scaling function is analytic for small arguments, i.e. $\phi(u) \simeq a_0 + a_1 u + \mathcal{O}(u^2)$, then the leading term in the expansion of the spectral scaling function will be proportional to $z^{-2}$, as the first term in Eq.~\eqref{scaling-function-long-time} vanishes for $\lambda = 0$.
As a result, the nonstationary spectral density, $S_\text{ta}(\omega,T) \propto T^{\alpha-1} \omega^{-2}$, exhibits a $\omega^{-2}$ frequency dependence like a normal diffusive process, which is recovered for $\alpha = 1$.
By contrast, if the expansion of $\phi(u)$ contains a nonanalytic term with $\lambda \not \in \mathbb{N}_0$, the corresponding term in the expansion of $\chi(z)$ proportional to $z^{-\lambda-1}$ will be non-zero. 
In this case,  to leading order, the spectral density behaves as $S_\text{ta}(\omega,T) \propto T^{\alpha-\lambda} \omega^{-\lambda-1}$.
This kind of frequency dependence is generally referred to as $1/f$-noise \cite{dut81}.
We will elaborate on this connection in more detail below.

\textit{Examples.}
Let us apply the above results to three  paradigmatic models of anomalous diffusion. 
These processes are characterized by a nonlinear increase of the mean-square displacement, $\langle x^2(t) \rangle = 2 D_\alpha t^{\alpha}$, where $D_\alpha$ is a generalized diffusion coefficient. Subdiffusion is obtained for $0 < \alpha < 1$ and superdiffusion for $\alpha > 1$ \cite{met00}.\\
\emph{(a) Scaled Brownian motion (SBM). }
SBM is maybe the simplest way to construct a stochastic process that exhibits anomalous diffusion.
It is defined as $x^\text{SBM}(t) = x(t^{\alpha})$, where $x(t)$ is a Brownian motion  \cite{lim02,thi14}.
Its autocorrelation function reads for $\tau > 0$,
\begin{align}
C^{\text{SBM}}(\tau,t) = 2 D_\alpha \text{min}(t^{\alpha},(t+\tau)^{\alpha}) = 2 D_\alpha t^{\alpha}, 
\end{align}
It is  of the scaling form \eqref{scaling-correlation} with $\mathcal{C}=2 D_\alpha$ and $\phi^\text{SBM}(u) = 1$.
SBM, like Brownian motion, is Markovian, however, its increments are nonstationary.\\
\emph{(b) Fractional Brownian motion (FBM).}
FBM is a non-Markovian generalization of Brownian motion.
One usually distinguishes between Riemann-Liouville (RL) FBM with nonstationary increments  \cite{lev53},
\begin{align}
x^\text{RL-FBM}(t) &= \int_{0}^{t} \text{d}x(t') (t-t')^{\frac{\alpha}{2}-1}, 
\end{align}
and Mandelbrot-van Ness (MN) FBM with stationary increments \cite{man68},
\begin{align}
x^\text{MN-FBM}(t) &= \int_{0}^{t} \text{d}x(t') (t-t')^{\frac{\alpha}{2}-1}  \\
&\quad + \int_{-\infty}^{0} \text{d}x(t') (t-t')^{\frac{\alpha}{2}-1} - (-t')^{\frac{\alpha}{2}-1}.\nonumber
\end{align}
The corresponding scaling autocorrelation functions are given by Eq.~\eqref{scaling-correlation} with \cite{lim02},
\begin{align}
\phi^{\text{RL-FBM}}(u) &= \alpha \int_{0}^{1} \text{d}w \ w^{\frac{\alpha-1}{2}} \big(w+u\big)^{\frac{\alpha-1}{2}}  \\
\phi^{\text{MN-FBM}}(u) &= \frac{1}{2}\big[1 +  \big(1+u\big)^{\alpha} - u^{\alpha} \big].
\end{align}
SBM and FBM motion play an important role in describing anomalous diffusion in biological living cells \cite{gol06,web10,jeo11,tab13}.\\
\emph{(c) Diffusion in a logarithmic potential (LOG). }
A Brownian particle diffusing in an asymptotically logarithmic potential of the form $U(x) \simeq U_0 \ln\vert x/a\vert$, for $x \gg a$, with free diffusion coefficient $D^{*}$  exhibits subdiffusive behavior for $1 < U_0/D^{*} < 3$, with a diffusion exponent $\alpha = 3/2 - U_0/2 D^{*}$ \cite{dec11}.
The autocorrelation function is also of the scaling form \eqref{scaling-correlation} with  a scaling function \cite{dec12a},
\begin{align}
&\phi^{\text{LOG}}(u) = \frac{\sqrt{\pi} \alpha}{\Gamma(3-\alpha)} u^{\alpha}  \\
& \quad \times \int_{0}^{\infty} \text{d}w \ w^2 e^{-w^2} {}_1 F_{1} \Big(\frac{3}{2},3-\alpha,w^2\Big) \Gamma\big(2-\alpha,w^2 u\big) . \nonumber
\end{align}
Diffusion in logarithmic potentials appears in a great variety of problems \cite{lut13}, ranging from DNA denaturization \cite{fog07}, systems with long-range interactions \cite{bou05} and diffusion of atoms in optical lattices \cite{lut03}.

\begin{figure}
\includegraphics[width=0.48\textwidth]{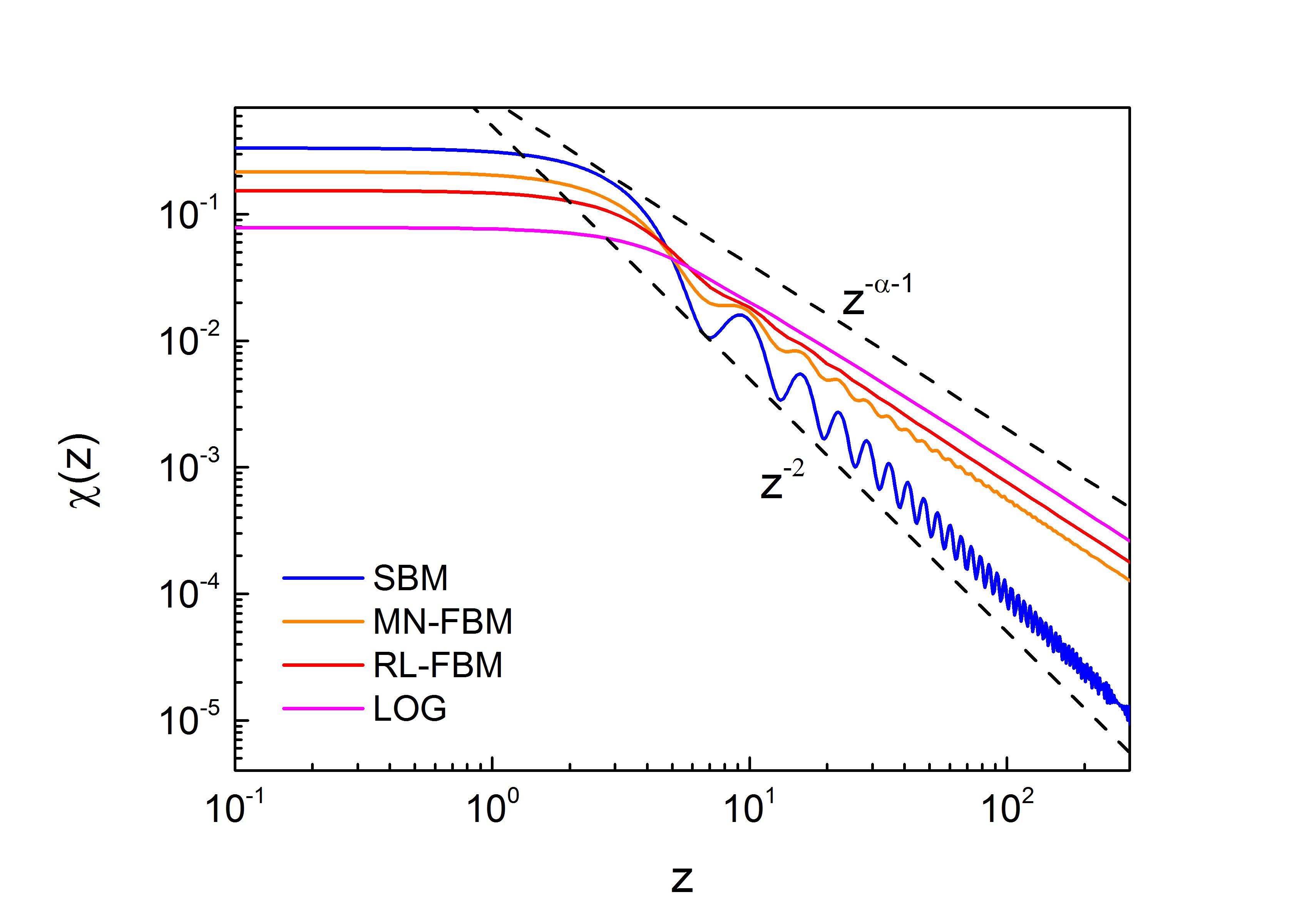}
\caption{(color online) Frequency scaling function $\chi(z)$ \eqref{scaling-wiener-Khinchin} for four models of anomalous diffusion: scaled Brownian motion (SBM), Riemann-Liouville (RL) and Mandelbrot-van Ness (MN) fractional Brownian motion (FBM) and diffusion in a logarithmic potential (LOG) for the scaling exponent $\alpha = 0.3$. For identical mean-square displacements in all cases, the asymptotic dependence  of $\chi(z)$  for large $z$ (dashed lines) is different, owing to the nonanalytic behavior of the time scaling function $\phi(y)$ for both types of FBM and LOG. The results shown here are exact results obtained by numerical integration of Eq.~\eqref{scaling-wiener-Khinchin}, the asymptotic expansion for large $z$ yields Eqs.~\eqref{spectral-density-SBM} through \eqref{spectral-density-superdiffusive}.}
\label{fig:scaling-function-frequency}
\end{figure}

\begin{figure}
\includegraphics[width=0.48\textwidth]{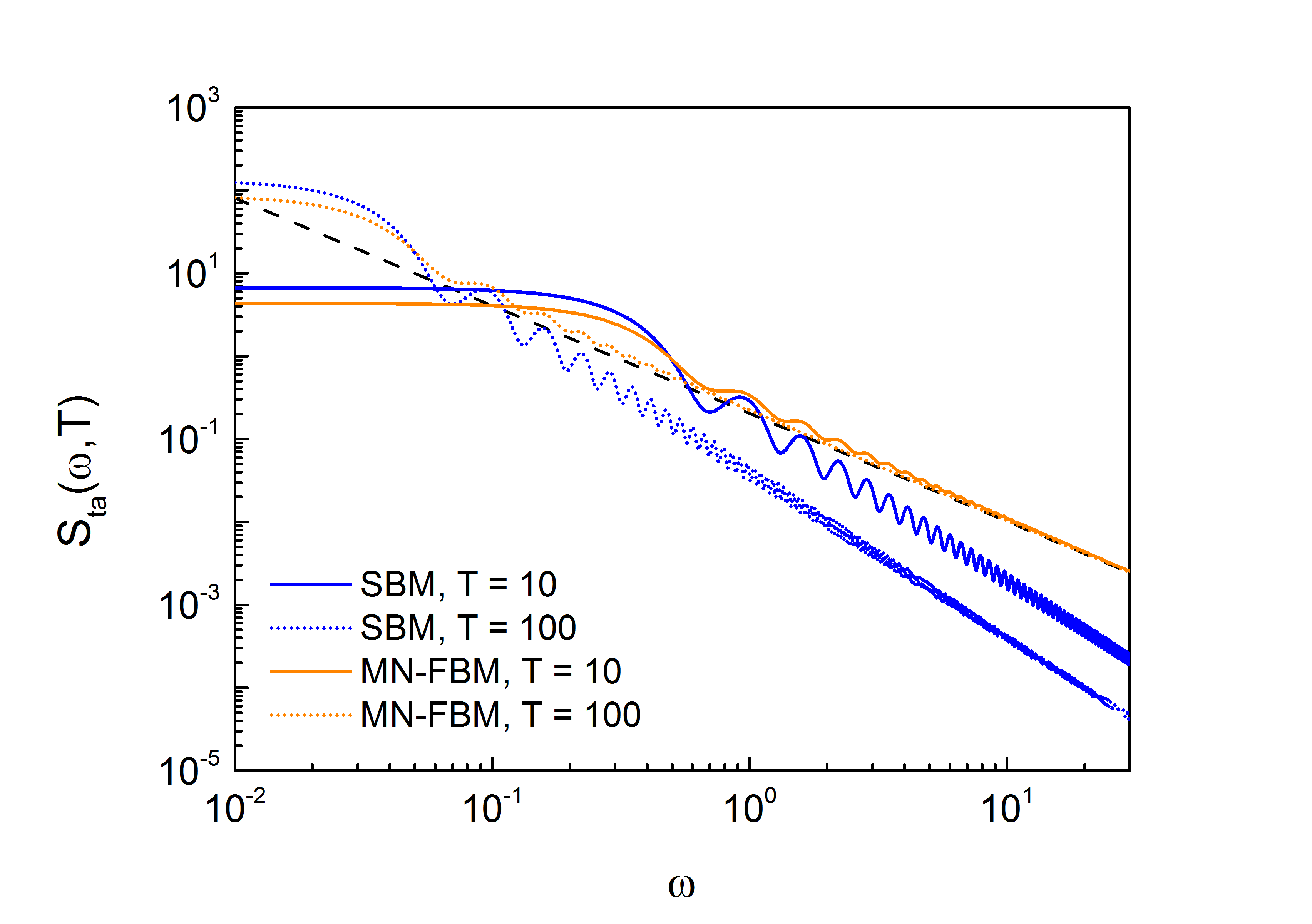}
\caption{(color online) Time-averaged spectral density $S_\text{ta}(\omega,T)$  \eqref{scaling-wiener-Khinchin} for scaled Brownian motion (SBM) and Mandelbrot-van Ness fractional Brownian motion (MN-FBM) for $\alpha = 0.3$ and different measurement times $T$. The spectral density converges to a stationary expression in the long-time limit for FBM, whereas for SBM, the magnitude of the high-frequency spectrum decreases with time.}
\label{fig:spectral-density}
\end{figure}

In the following, we compute the  spectral density \eqref{5} for the three processes by applying the nonstationary Wiener-Khinchin theorem (12).
For SBM, the scaling function is unity and thus analytic.
By Eq.~\eqref{scaling-function-long-time}, we then have $\chi(z) \propto z^{-2}$ for large $z$ (see Fig.~\ref{fig:scaling-function-frequency}).
Consequently, the spectral density is for $\omega T \gg 1$,
\begin{align}
S^\text{SBM}_\text{ta}(\omega,T) &\simeq 4 D_\alpha \omega^{-2} T^{\alpha-1} \label{spectral-density-SBM},
\end{align}
with an algebraic dependence on the measurement time $T$. If the process is subdiffusive, $0<\alpha<1$, the magnitude of the high-frequency spectrum decreases with time, as shown in Fig.~\ref{fig:spectral-density}.
The rescaling $t \rightarrow t^{\alpha}$ in SBM can thus be understood as a continuous slowing down of the process, which reduces the prevalence of high-frequency components in the spectrum.
In the superdiffusive regime, $1<\alpha<2$, the process speeds up and the magnitude of the high-frequency spectrum increases with time.
In both cases, the dependence on frequency is precisely the same as for usual Brownian motion.

On the other hand, for both kinds of FBM and for the logarithmic potential,  we have $\phi(u) \simeq a_0 + a_{\alpha} u^{\alpha} + a_1 u$ (the explicit expressions for the coefficients are given in the Supplementary Material \cite{sm}), and hence $\chi(z) \propto z^{-\alpha-1}$ for $\alpha < 1$, see Fig.~\ref{fig:scaling-function-frequency}.
The time-averaged spectral density in the subdiffusive regime, $0<\alpha<1$,  and for long measurement times,  $\omega T \gg 1$,  is to leading order given by,
\begin{align}
S_\text{ta}^\text{FBM,LOG}(\omega) &\simeq 2 \mathcal{C} a_{\alpha} \Gamma(\alpha+1) \sin\Big(\frac{\pi \alpha}{2}\Big) \omega^{-\alpha-1} . \label{spectral-density-stationary}
\end{align}
Interestingly, even though the process  is nonstationary, the spectral density is  asymptotically independent of the measurement time, see Fig.~\ref{fig:spectral-density}.
This shows that a stationary spectral density does not necessarily imply a stationary process, the nonstationarity being in this case encoded in the low-frequency cutoff.
Equation \eqref{scaling-function-long-time} also allows  to evaluate finite-time corrections to leading order,
\begin{align}
S_\text{ta}^\text{FBM,LOG}(\omega,T) &\simeq 2 \mathcal{C} \omega^{-\alpha-1} \bigg[ a_{\alpha} \Gamma(\alpha+1) \sin\Big(\frac{\pi \alpha}{2}\Big) \nonumber \\
&\quad + a_{\alpha} \Gamma(\alpha+1) \cos\Big(\frac{\pi \alpha}{2}\Big) (\omega T)^{-1} \nonumber \\
&\quad + \Big(a_0 - \frac{a_1}{\alpha}\Big) (\omega T)^{\alpha-1} \bigg] ,\label{spectral-density-stationary-corrections}
\end{align}
These corrections have two different origins:
The dominant terms proportional to $a_0$ and $a_1$ stem from the nonstationarity  of the correlation function.
The correction term proportional to $a_{\alpha}$, on the contrary, appears even if the system is perfectly stationary (e.g. for $\alpha < 0$) and is due to the finite measurement time.
In the superdiffusive regime, $1 <\alpha < 2$, the spectral density of FBM also displays an algebraic time-dependence to leading order,
\begin{align}
S_\text{ta}^\text{FBM}(\omega,T) &\simeq 2 \mathcal{C} \Big(a_0 - \frac{a_1}{\alpha} \Big) \omega^{-2} T^{\alpha-1} \label{spectral-density-superdiffusive}.
\end{align}
The overall magnitude of the spectral density increases with the measurement time $T$.
This is similar to the result \eqref{spectral-density-SBM} for superdiffusive SBM.

\textit{Relation to $1/f$-noise.} As we have seen in Eq.~\eqref{spectral-density-stationary}, both kinds of FBM and the diffusion in a logarithmic potential lead to a spectral density of the form $S_\text{ta}(\omega) \propto \omega^{-\alpha-1}$.
Generally, a spectrum with $S(\omega) \propto \omega^{-\beta}$ with $0<\beta<2$ is termed $1/f$-noise \cite{dut81}.
This kind of spectrum was first observed in flicker noise in triodes \cite{joh25}, and has since been found in a wide range of other systems in physics \cite{wei88}, biology \cite{gil95}, geology \cite{dav02} and  phonology \cite{vos75}.
The scaling Wiener-Khinchin relation \eqref{scaling-wiener-Khinchin} shows that $1/f$-noise naturally occurs for processes with scale-invariant correlations.
The main requirement is that the correlation function  is nonanalytic in the limit $t \gg \tau$.
Out of the three examples we investigated above, this is the case for FBM and the diffusion in a logarithmic potential, with the former being non-Markovian but linear, and the latter  Markovian but nonlinear.
For $\alpha < 0$, these processes are stationary and the power spectrum, which is proportional to $\omega^{-\alpha-1}$, is integrable at low frequencies, yielding a finite stationary power.
For $0 < \alpha < 1$, by contrast, the stationary spectrum is nonintegrable at low frequencies (infrared catastrophe), formally leading to infinite power.
This has caused some discussion on the physical nature of this divergence \cite{man67}.
Using the scaling Wiener-Khinchin relation \eqref{scaling-wiener-Khinchin}, this apparent contradiction is easily resolved, since for a diffusive process, the power actually diverges in the infinite time limit.
For any finite time, there is a low-frequency cutoff on the spectrum that depends on the measurement time, which ensures finite power.
As the measurement time increases, this cutoff moves to lower and lower frequencies, and the process approaches $1/f$-noise at all frequencies.
We note that a similar behavior was recently observed for another scale invariant (superdiffusive) stochastic process, the L{\'e}vy walk, in application to blinking quantum dots \cite{nie13,sad14}. Our approach allows to generalize these results to generic subdiffusive random processes.

\textit{Conclusion.} The generalized Wiener-Khinchin theorem \eqref{spectral-nonstationary}-\eqref{spectral-nonstationary-inverse} extends  the notion of a power spectrum to nonstationary stochastic  processes in a natural way and establishes a direct connection to their autocorrelation functions. By studying three ubiquitous scale-invariant models of anomalous diffusion, we have demonstrated its usefulness in analyzing the properties of aging processes. Our results thus provide an extension of the range of applicability of spectral analysis---in terms of a proper (positive) spectral density---to the nonstationary regime.

\begin{widetext}

\section*{Supplemental material}

\renewcommand{\theequation}{S\arabic{equation}}
\makeatletter
\renewcommand{\thefigure}{S\@arabic\c@figure}
\makeatother

\subsection*{Large argument expansion of $\chi(z)$}
In order to obtain the long-time limit of the spectral density, we need to know the asymptotic behavior of the function $\chi(z)$ for large arguments.
From Eq.~(12) of the main text, $\chi(z)$ is defined as,
\begin{align}
\chi(z) = \int_{0}^{1} \text{d}v \ v^{\alpha+1} \cos(v z) \int_{\frac{v}{1-v}}^{\infty} \text{d}u \ u^{-\alpha-2} \phi(u)
\end{align}
We define the auxiliary function,
\begin{align}
f(v) \equiv v^{\alpha+1} \int_{\frac{v}{1-v}}^{\infty} \text{d}u \ u^{-\alpha-2} \phi(u) .
\end{align}
For the scaling function $\phi(u)$, we assume the asymptotic behavior,
\begin{align}
\phi(u) \simeq \left\lbrace \begin{array}{ll}
a_\lambda u^{\lambda} &\text{for} \quad u \ll 1 \\
b_\mu u^{\mu} &\text{for} \quad u \gg 1,
\end{array} \right.
\end{align}
with $\lambda > -1$ and $\mu < \alpha + 1$. Using this, it is easy to show that,
\begin{align}
f(v) \simeq \left\lbrace  \begin{array}{ll}
&a_\lambda \Big[ \frac{1}{\alpha+1-\lambda} v^{\lambda} - v^{\lambda+1}\Big] \quad \text{for} \; v \ll 1 \\[2 ex]
&b_\mu \Big[ \frac{1}{\alpha+1-\mu} (1-v)^{\alpha+1-\mu} - \frac{\mu}{\alpha+1-\mu} (1-v)^{\alpha+2-\mu} \Big] \quad \text{for} \; 1-v \ll 1 . \label{f-u-expansion}
\end{array} \right.
\end{align}
We now split the integral into three parts,
\begin{align}
\bigg[ \int_{0}^{\epsilon} + \int_{\epsilon}^{1-\epsilon} + \int_{1-\epsilon}^{1} \bigg] \text{d}v \ \cos(v z) f(v) ,
\end{align}
where $\epsilon$ is small enough that the expansion \eqref{f-u-expansion} is justified.
Using this expansion, we find for large $z$,
\begin{align}
\int_{0}^{\epsilon} \text{d}v \ \cos(v z) f(v) &\simeq a_\lambda \Bigg[ \frac{1}{\alpha+1-\lambda} \bigg[ -\Gamma(\lambda+1) \sin\Big(\frac{\pi \lambda}{2}\Big) z^{-\lambda-1} + \frac{\epsilon^{\lambda} \sin(z \epsilon)}{z} + \frac{\lambda\epsilon^{\lambda-1}\cos(z \epsilon)}{z^2} \bigg] \nonumber \\
& \; - \bigg[ -\Gamma(\lambda+2) \cos\Big(\frac{\pi \lambda}{2}\Big) z^{-\lambda-2} + \frac{\epsilon^{\lambda+1} \sin(z \epsilon)}{z} +  \frac{(\lambda+1)\epsilon^{\lambda}\cos(z \epsilon)}{z^2} \bigg] \Bigg] \nonumber \\
\int_{1-\epsilon}^{1} \text{d}v \ \cos(v z) f(v) &\simeq b_\mu \Bigg[ \frac{1}{\alpha+1-\mu} \bigg[ -\Gamma(\alpha+2-\mu) \cos\Big(z - \frac{\pi (\alpha-\mu)}{2}\Big) z^{\mu-\alpha-2} - \frac{\epsilon^{\alpha+1-\mu} \sin(z (1-\epsilon))}{z} \nonumber \\
&\qquad + \frac{(\alpha+1-\mu)\epsilon^{\alpha-\mu}\cos(z (1-\epsilon))}{z^2} \bigg] \nonumber \\
& \; - \frac{\mu}{\alpha+1-\mu} \bigg[ \Gamma(\alpha+3-\mu) \sin\Big(z - \frac{\pi (\alpha-\mu)}{2}\Big) z^{\mu-\alpha-3} - \frac{\epsilon^{\alpha+2-\mu} \sin(z (1-\epsilon))}{z} \nonumber \\
&\qquad + \frac{(\alpha+2-\mu)\epsilon^{\alpha-\mu}\cos(z (1-\epsilon))}{z^2} \bigg] \Bigg]
\end{align}
In the center part of the integral, we may integrate by parts,
\begin{align}
\int_{\epsilon}^{1-\epsilon}&\text{d}v \ \cos(v z) f(v) = \frac{\sin(v z)}{z} f(v) \bigg\vert_{\epsilon}^{1-\epsilon} + \frac{\cos(v z)}{z^2} f'(v) \bigg\vert_{\epsilon}^{1-\epsilon} + \mathcal{O}(z^{-3}) .
\end{align}
Again using Eq.~\eqref{f-u-expansion} for the boundary terms, we see that all terms depending on $\epsilon$ cancel and we are left with,
\begin{align}
\chi(z) &\simeq a_\lambda \bigg[ - \frac{\Gamma(\lambda+1)}{\alpha+1-\lambda} \sin\Big(\frac{\pi \lambda}{2}\Big) z^{-\lambda-1} + \Gamma(\lambda+2) \cos\Big(\frac{\pi \lambda}{2}\Big) z^{-\lambda-2} \bigg] \nonumber \\
& \quad + \frac{b_\mu}{\alpha+1-\mu} \bigg[ -\Gamma(\alpha+2-\mu) \cos\Big(z - \frac{\pi (\alpha-\mu)}{2}\Big) z^{\mu-\alpha-2} - \mu \Gamma(\alpha+3-\mu) \sin\Big(z - \frac{\pi (\alpha-\mu)}{2}\Big) z^{\mu-\alpha-3} .
\end{align}
For $\mu < \alpha - \lambda$, which is the case for all the examples discussed in the main text, the leading order expression is given by the part proportional to $a_\lambda$, which is precisely Eq.~(14) of the main text.

\subsection*{Asymptotic expressions for $\phi(u)$ for the example systems}

\begin{table}[h!]
\begin{tabular}{|c|c|c|c|c|}
\hline 
{} & $\lbrace \lambda_{i} \rbrace$ & $ \lbrace a_{\lambda_i} \rbrace$ & $\lbrace \mu_i \rbrace$ & $\lbrace b_{\mu_i} \rbrace$ \\ 
\hline 
SBM & $0$ & $1$ & $0$ & $1$ \\ 
\hline 
RL-FBM & $0; \ \alpha; \ 1; \ \ldots$ & $1; \ -\frac{\sqrt{\pi} \Gamma((\alpha+1)/2)}{\Gamma(\alpha/2) \sin(\pi\alpha/2)}; \ \frac{\alpha}{2}; \ \ldots$ & $\frac{\alpha-1}{2}; \ \frac{\alpha-3}{2}; \ \frac{\alpha-5}{2}; \ \ldots$ & $\frac{2\alpha}{\alpha+1}; \ \frac{\alpha(\alpha-1)}{\alpha+3}; \ \frac{\alpha(\alpha-1)(\alpha-3)}{4(\alpha+5)}; \ \ldots$ \\ 
\hline 
MN-FBM & $0; \ \alpha; \ 1; \ \ldots$ & $1; \ -1; \ \frac{\alpha}{2}; \ \ldots$ & $0; \ \alpha-1; \ \alpha-2; \ \ldots$ & $\frac{1}{2}; \ \frac{\alpha}{2}; \ \frac{\alpha(\alpha-1)}{4}; \ \ldots$ \\ 
\hline 
LOG & $0; \ \alpha; \ \ldots$ & $\frac{\sqrt{\pi}}{2 \Gamma(3-\alpha)}; \ -\frac{\pi \Gamma(2-\alpha) \Gamma(1-\alpha)}{4 \Gamma^2(3/2-\alpha)}; \ \ldots$ & $\alpha-\frac{3}{2}; \ \alpha-\frac{5}{2}; \ \ldots$ & $\frac{\sqrt{\pi} \alpha \Gamma(7/2-\alpha)}{3 \Gamma(3-\alpha)}; \ \frac{\sqrt{\pi} \alpha (3-2\alpha) \Gamma(9/2-\alpha)}{10(\alpha-3) \Gamma(3-\alpha)}; \ \ldots$\\ 
\hline 
\end{tabular} 
\end{table}

In the main text, we discuss the spectral density for scaled Brownian motion (SBM), Riemann-Liouville (RL-FBM) and Mandelbrot-van Ness (MN-FBM) fractional Brownian motion and diffusion in a logarithmic potential (LOG).
The table lists the explicit expressions for the coefficients and exponents of the small argument ($\phi(u) \simeq a_{\lambda_1} u^{\lambda_1} + a_{\lambda_2} u^{\lambda_2} + \ldots$) and large argument ($\phi(u) \simeq b_{\mu_1} u^{\mu_1} + b_{\mu_2} u^{\mu_2} + \ldots$) expansions of the corresponding scaling functions.
The first column lists the stochastic model, the second column the first few exponents $\lambda_i$ in the small-argument expansion in ascending order (for $0 < \alpha < 1$).
The third column lists the coefficients corresponding to these exponents.
The fourth and fifth column are the same, but for the large-argument expansion.
So, for example, the first three terms in the small argument expansion for the scaling function for MN-FBM motion are,
\begin{align}
\phi(u) \simeq 1 - u^{\alpha} + \frac{\alpha}{2} u + \mathcal{O}(u^2).
\end{align}
Note that each of the exponents $\lambda_i$ and $\mu_i$ individually yields a contribution of the type (S8) to the expansion of the frequency scaling function $\chi(z)$.

\end{widetext}

\end{document}